\documentclass[superscriptaddress,twocolumn,amsmath,amssymb,aps,prb]{revtex4-1}

\usepackage{graphicx}
\usepackage{dcolumn}
\usepackage{bm}
\usepackage{natbib}

\usepackage[breaklinks=true,colorlinks=true,linkcolor=blue,urlcolor=blue,citecolor=blue]{hyperref}

\begin{document}
\title{The occupied electronic structure of ultrathin boron doped diamond}

\author{A.~K.~Schenk}
\affiliation{Center for Quantum Spintronics, Department of Physics, Norwegian University of Science and Technology, NO-7491 Trondheim, Norway}
\author{A.~C.~Pakpour-Tabrizi}
\affiliation{London Centre for Nanotechnology and Department of Electronic and Electrical Engineering, University College London, 17-19 Gordon Street, London WC1H 0AH, U.K.}
\author{A.~J.~U.~Holt}
\author{S.~K.~Mahatha}
\author{F.~Arnold}
\author{M.~Bianchi}
\affiliation{Department of Physics and Astronomy, Interdisciplinary Nanoscience Center, Aarhus University, 8000 Aarhus C, Denmark}
\author{R.~B.~Jackman}
\affiliation{London Centre for Nanotechnology and Department of Electronic and Electrical Engineering, University College London, 17-19 Gordon Street, London WC1H 0AH, U.K.}
\author{J.~A.~Miwa}
\author{Ph.~Hofmann}
\affiliation{Department of Physics and Astronomy, Interdisciplinary Nanoscience Center, Aarhus University, 8000 Aarhus C, Denmark}
\author{S. P. Cooil}
\affiliation{Center for Quantum Spintronics, Department of Physics, Norwegian University of Science and Technology, NO-7491 Trondheim, Norway}
\affiliation{Department of Physics, Aberystwyth University, Aberystwyth SY23 3BZ, United Kingdom}
\author{J.~W.~Wells}
\altaffiliation[justin.wells@ntnu.no]{}
\affiliation{Center for Quantum Spintronics, Department of Physics, Norwegian University of Science and Technology, NO-7491 Trondheim, Norway}
\author{F.~Mazzola}\altaffiliation[Present Address: ]{SUPA, School of Physics and Astronomy, University of St. Andrews, St. Andrews KY16 9SS, UK}
\affiliation{Center for Quantum Spintronics, Department of Physics, Norwegian University of Science and Technology, NO-7491 Trondheim, Norway}

\begin{abstract}
Using angle-resolved photoelectron spectroscopy, we compare the electronic band structure of an ultrathin (1.8~nm) $\delta$-layer of boron-doped diamond with a bulk-like boron doped diamond film ($\rm{3~\mu m}$). Surprisingly, the measurements indicate that except for a small change in the effective mass, there is no significant difference between the electronic structure of these samples, irrespective of their physical dimensionality. While this suggests that, at the current time, it is not possible to fabricate boron-doped diamond structures with quantum properties, it also means that nanoscale doped diamond structures can be fabricated which retain the classical electronic properties of bulk-doped diamond, without a need to consider the influence of quantum confinement.
\end{abstract}

\pacs{Valid PACS appear here}%

\maketitle
Diamond is an electrical insulator with spectacular physical properties: it is one of the hardest natural materials,\cite{Thompson:2000} has one of the highest thermal conductivities of any elemental material,\cite{Berman:1953,Thompson:2000} a high breakdown field, biocompatibility\cite{Grill:2003} and, contrary to traditional semiconductors, is robust against radiation damage.\cite{Bauer:1995} Diamond may be doped with boron either naturally, or during Chemical Vapour Deposition (CVD) film growth\cite{Spitsyn:1981} or with post-growth ion implantation,\cite{Braunstein:1983} turning diamond semiconducting,\cite{Collins:1971,Chrenko:1973} metallic or allowing a superconducting transition under the right conditions,\cite{Ekimov:2004,Yokoya:2005, Bustarret:2004} depending on the dopant concentration. These properties make diamond an appealing candidate for a variety of electronic applications.\cite{Wort:2008,Kalish:2007} Growing ultrathin (nanometer scale) diamond films may allow minituarised devices to benefit from the exemplary properties of diamond, as well as reducing processing costs for applications where only a thin film is required. In recent years, the ability to grow ultrathin, heavily boron doped diamond layers has been demonstrated\cite{Butler:2017,Vikharev:2016,Volpe:2012} -- such doped profiles are typically referred to as $\delta$-doping (or $\delta$-layers) and may have strongly modified electronic properties when compared to thicker films.\cite{Miwa:2013,Miwa:2014a,Mazzola:2014a,Mazzola:2018,Sullivan:2004}

$\delta$-doping consists of engineering a narrow profile (typically from one atomic layer to several nanometers) of electron donor or acceptor species within a host material, either submerged in the bulk (encapsulated, or so-called ``capped'' $\delta$-layers) or at the surface (unencapsulated, ``uncapped'' $\delta$-layers),\cite{Gossmann:1993,Harris:1993,Schubert:1996} such that the layer thickness is narrow relative to the ground state wavefunction of the free carrier gas.\cite{Schubert:1990} These structures have electronic properties dictated by the interplay of quantum confinement effects, spin\cite{Menshov:2009} and charge ordering and the overlap between the host material and the dopants' atomic-wavefunctions.\cite{Gossmann:1993,Schubert:1996} As an example, phosphorus doped $\delta$-layers in silicon (referred to as Si:P $\delta$-layers) create new low-dimensional electronic states\cite{Miwa:2013,Miwa:2014a,Mazzola:2014a} which influence electrical transport properties.\cite{Sullivan:2004} The self-consistent Poisson-Schr{\"o}dinger calculations of Chicot~\textit{et al.}\ \cite{Chicot:2014} and Fiori~\textit{et al.}\ \cite{Fiori:2010} indicate that boron-doped $\delta$-layers in diamond, with experimentally achievable thicknesses and dopant densities, will generate a potential which is sufficiently strong and narrow to create quantum confined states; however these states, or any other alterations to the electronic structure as a result of quantum confinement, are yet to be experimentally confirmed.

\begin{figure}
	\centering
    \includegraphics[width=0.7\columnwidth]{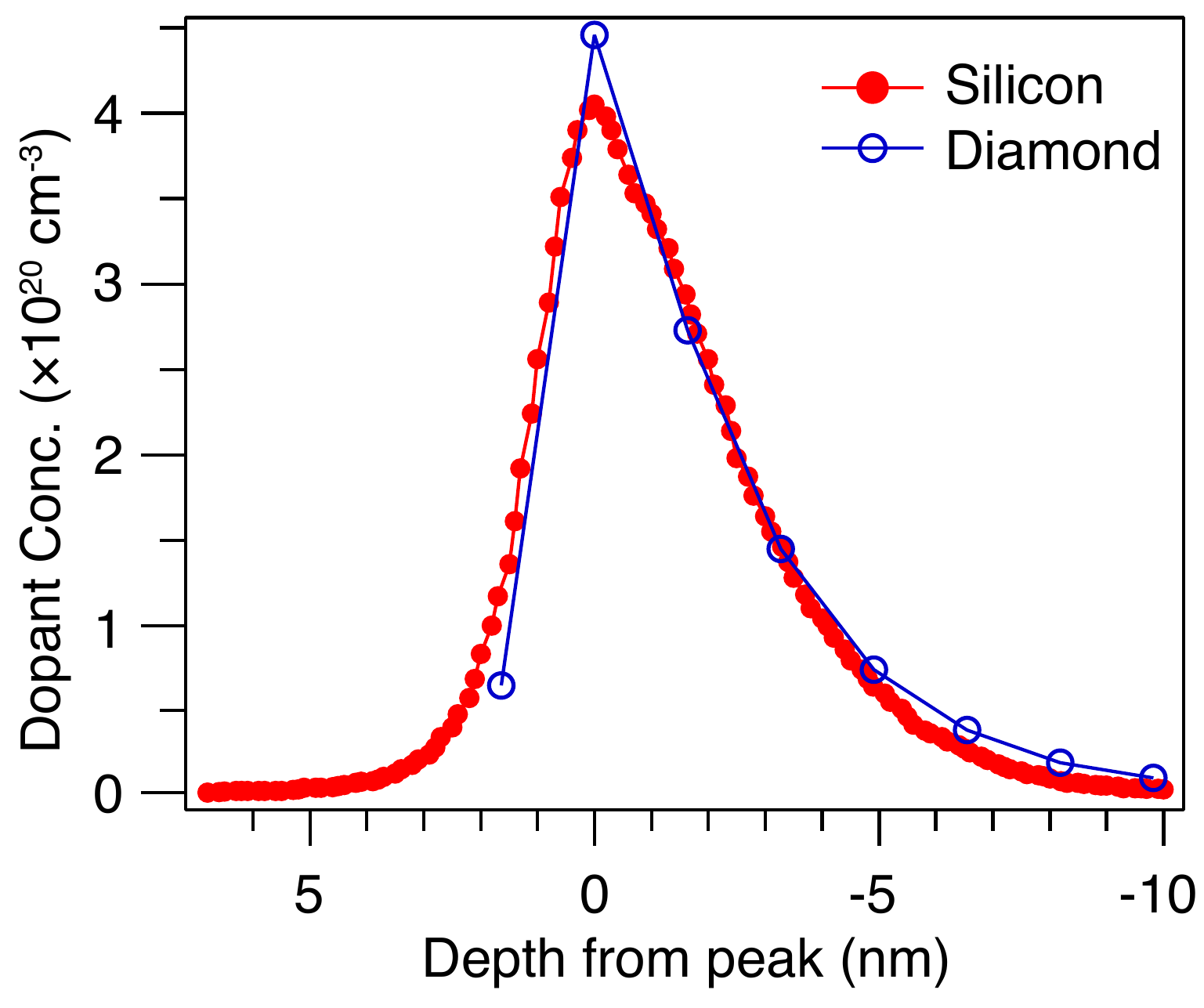}
    \caption{\textbf{SIMS depth profile of a $\delta$ doped silcon and diamond sample.} The measured dopant concentration in a boron-doped diamond $\delta$-layer with nominal thickness 1.8~nm is compared with a phosphorus-doped silicon $\delta$-layer of nominal thickness 2~nm.}
	\label{fig:SIMS}
\end{figure}

In this study, we use angle-resolved photoelectron spectroscopy (ARPES) to compare the electronic structure of a 1.8~nm boron doped $\delta$-layer with that of a thick $3~\mu\rm{m}$ boron-doped diamond film, to explore if the electronic structure is modified by nanoscale confinement. ARPES has been demonstrated as an exemplary tool for characterising the occupied electronic structure of low-dimensional systems, providing a clear and unique means of distinguishing electronic states associated with reduced dimensionality from three-dimensional electronic structure,\cite{Damascelli:2004,Himpsel:1980} and has been successfully applied to numerous investigations of Si:P $\delta$-layers,\cite{Miwa:2013,Miwa:2014a,Mazzola:2014a,Mazzola:2018} where new states are formed due to the quantum confinement. Contrary to expectation, our results indicate that the electronic structure of currently achievable $\delta$-doped diamond films is very similar to that of bulk doped diamond. This finding offers some explanation for the observed lack of quantum confinement enhancement in the transport measurements performed by Chicot~\textit{et al.}\,\cite{Chicot:2012,Chicot:2014} and we discuss possible sources of the consistent discrepancy between theoretical expectation and experimental observation. 



\section*{Experimental Details}
This study uses a boron-doped $\delta$-layer sample with a nominal thickness of 1.8~nm, and a thick (thickness~$\sim\rm{3~\mu m}$) boron-doped film. The boron doped $\delta$-layer was grown on a $\rm{3.6~mm \times 3.6~mm}$ (100) oriented high pressure high temperature (HPHT) Ib substrate, with an intrinsic buffer layer (nominal thickness $\rm{0.5~\mu m}$) grown using CVD between the $\delta$-layer and substrate. As a comparison, a boron doped thick film was also grown with CVD. The boron doping density, determined with Secondary-Ion Mass Spectrometry (SIMS) is similar in all samples ($\sim5\times10^{20}~\rm{cm}^{-3}$). For details, see Refs.\ \onlinecite{Butler:2017,Balmer:2013,supp}.

The \textit{in situ} sample preparation consisted of annealing to 350$^{\circ}$C for 8 hours to remove atmospheric contamination, followed by multiple 5 second flashes to 800$^{\circ}$C. All data has been acquired at room temperature, with the $k_\parallel$ axis aligned along the $\rm{X}-\Gamma-\rm{X}$ direction, determined from the symmetry of constant energy maps acquired during sample alignment. The $k_\perp$ axis is likewise along $\rm{X}-\Gamma-\rm{X}$. A free-electron final state model\cite{Himpsel:1980} with an inner potential of 22~eV\cite{Edmonds:2013} has been used for converting units of photon energy into $k_\perp$.\footnote{Many values for the diamond inner potential are found in the literature; Ashenford and Lisgarten find 18.2~eV using the Shinohara method,\cite{Ashenford:1983} while previous ARPES work at similar photon energies to our study find values between 17.7~eV\cite{Guyot:2015} and 23~eV.\cite{Yokoya:2005} For the photon energies range $380$-$460$~eV this range of values for the inner potential gives rise to a maximum 0.8\% error ($\pm 0.061$~\AA\textsuperscript{-1}) in the presented $k_\perp$ values.} The photon energy range used in this work is relatively high compared to the typical photon energies used for ARPES. This is necessary as a result of diamond displaying non free-electron final state behaviour in measurements performed at low photon energies;\cite{Himpsel:1980b} a discussion of this, with supporting data, is presented in Ref.~\onlinecite{supp}. 

Relative energy alignment between measurements has been performed by acquiring the Fermi edge of a gold foil in electrical contact with the sample and aligning this to a common origin for all photon energies in this study. An absolute energy calibration has been performed at $h\nu$=520~eV by integrating the photoemission background (away from any strong features) and identifying the Fermi level. This atypical second step is necessary to compensate for the possibility of a Schottky barrier between the sample and calibration foil,\cite{Teraji:2014} as well as the possibility of a photovoltage generated by synchrotron light exposure,\cite{Williams:2014,Bandis:1996} both of which will manifest as an offset in the energy scale of the dataset.   

\section*{Results and Discussion}

\begin{figure}
	\centering
	\includegraphics[width=\columnwidth]{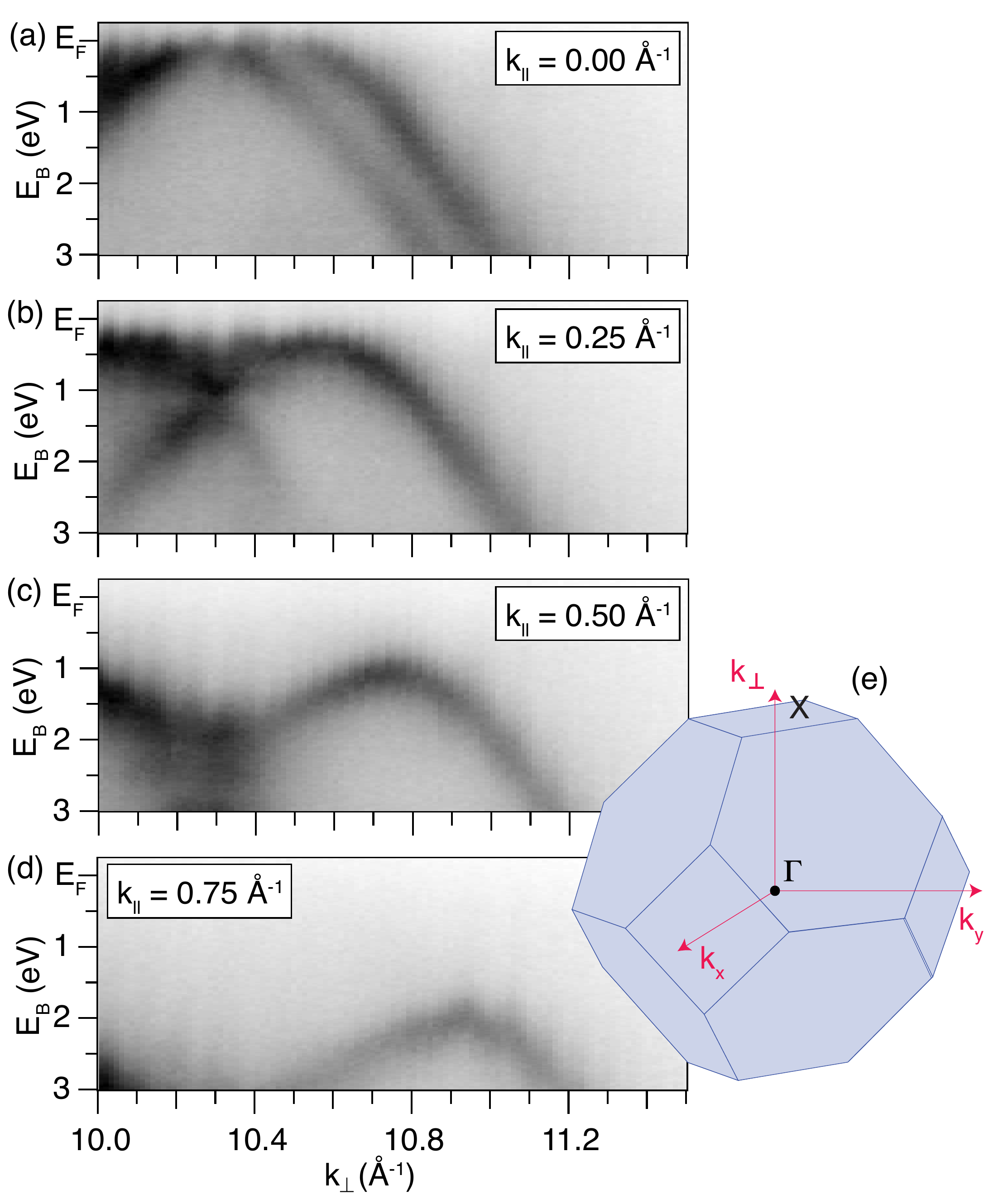}
	\caption{\textbf{Selected band dispersions with respect to $k_{\perp}$.} (a) - (d) Constant ${k}_{||}$ slices through the photon energy-dependent dataset ($380 - 460$~eV) acquired on the $\delta$-layer sample; the photon energy axis has been converted to $k_{\perp}$ using the assumption fo a free-electron-like final state ($k_{\perp}\approx$ 10.0 - 11.4 \AA$^{-1}$), and the values of ${k}_{||}$ chosen are shown in the panels. (e) Schematic of the bulk Brillouin Zone showing the definitions of the axes used.}
	\label{fig:Constkperp}
\end{figure}

Published calculations \cite{Chicot:2014,Fiori:2010} show the formation of a confining potential approximately 0.25~eV deep (for a 0.36~nm thick $\delta$-layer, and increasing with layer thickness), inducing confined hole states both above and below the Fermi level, with the typical characteristics expected of such quantum wells - the energy separation between states decreases for thicker $\delta$-layers, and decreases the closer the state is to the top of the well. These simulations assume an atomically sharp $\delta$-layer, with an immediate transition between the heavily doped $\delta$-layer and the surrounding diamond. In practice, boron $\delta$-layers in diamond are grown by adding a boron precursor to the diamond CVD growth process; such a growth process can yield a very sharp junction,\cite{Mer-Calfati:2014,Volpe:2012} but not as abrupt as in the calculations.  In order to address this possible inconsistency, we have carried out SIMS measurements (Fig.~\ref{fig:SIMS}) on both the diamond sample used here and a Si:P $\delta$-layer with a similar nominal thickness, used in previous work \cite{Polley:2013a}, and where pulsed laser atom probe tomography (PLAPT) was also used to confirm the sharpness of the profile. In both cases, the peak doping density, sharpness and width are extremely similar. However, it is also worth pointing out that the SIMS data presented is resolution limited, and hence it is possible that the profile is significantly sharper than Fig.~\ref{fig:SIMS} appears to indicate.  In any case, based on our previous work on Si:P $\delta$-layers \cite{Miwa:2013,Miwa:2014a,Mazzola:2014a,Mazzola:2018}, we expect such a dopant profile in diamond to give rise to strongly confined 2-dimensional quantum-well states.

For a system possessing states as a result of quantum confinement in the direction perpendicular to the surface, one expects to observe features which do not disperse with $k_{\perp}$.  Therefore, slices of constant $k_{\parallel}$ slices have been extracted from the ARPES dataset acquired on the $\delta$-layer sample, and are presented in Fig.~\ref{fig:Constkperp}. Within this representation of the data confined states will be present as non-dospersing features (i.e.\ horizontal lines across the panels in Fig.~\ref{fig:Constkperp}), with a varying intensity due to the changing photoemission transition matrix elements.\cite{Moser:2017,Louie:1980} Fig.~\ref{fig:Constkperp} shows no such horizontal features at any value of $k_\parallel$, suggesting that there are no occupied states uniquely associated with reduced electronic dimensionality within the $\delta$-layer.

\begin{figure}
	\centering
	\includegraphics[width=0.8\columnwidth]{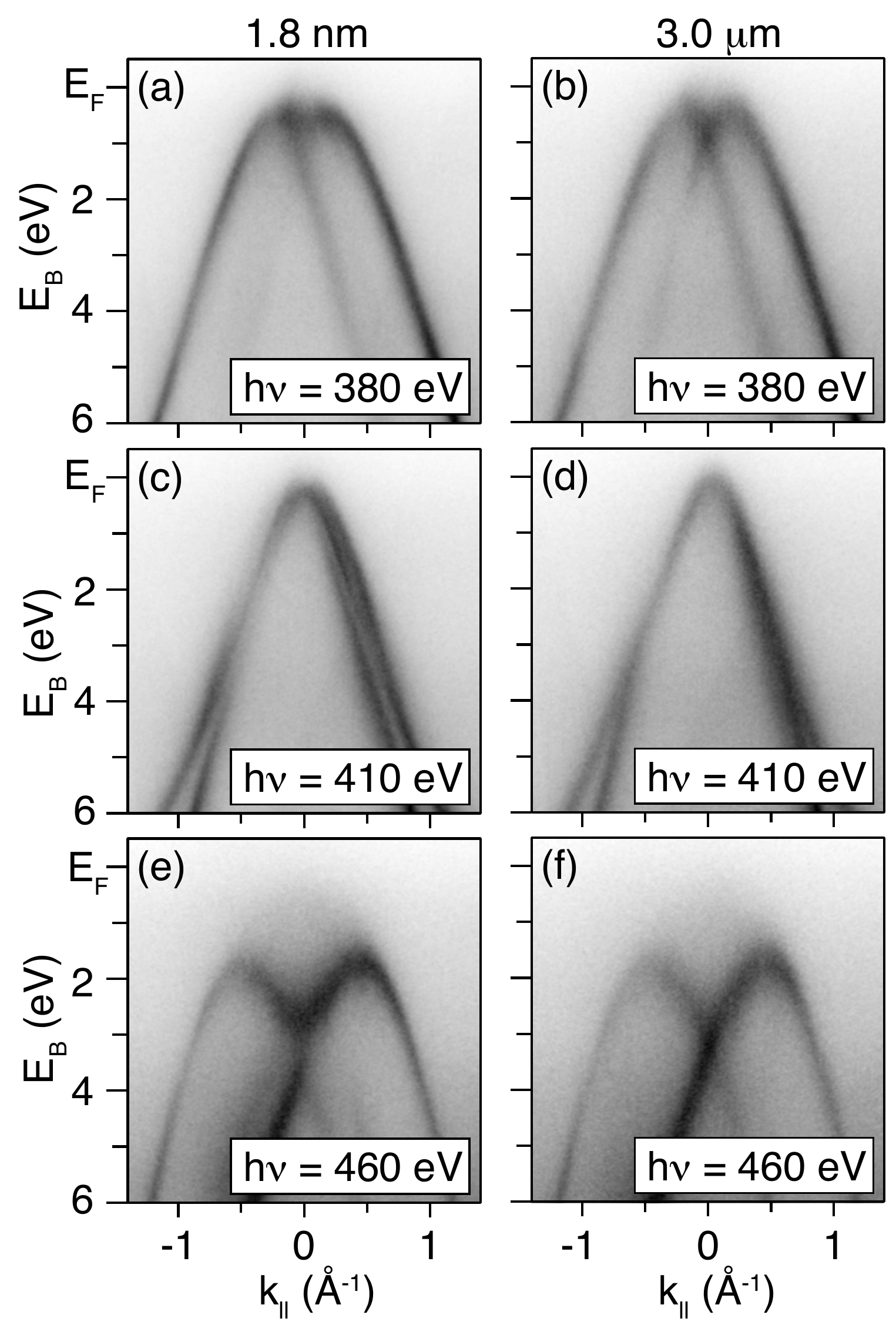}
	\caption{\textbf{Selected ARPES measurements performed on a 1.8~nm diamond $\delta$-layer sample and the $3.0~\mu$m bulk sample.} Measurements performed at: (a) and (b), a photon energy of 380~eV ($\approx$ k$_\perp$=10.0), (c) and (d), a photon energy of 410~eV ($\approx$ k$_\perp$10.3, corresponding to the bulk BZ center) and (e) and (f), a photon energy of 460~eV ($\approx$ k$_\perp$11.5).}
	\label{fig:Ekpara}
\end{figure}

\begin{figure}
	\centering
	\includegraphics[width=\columnwidth]{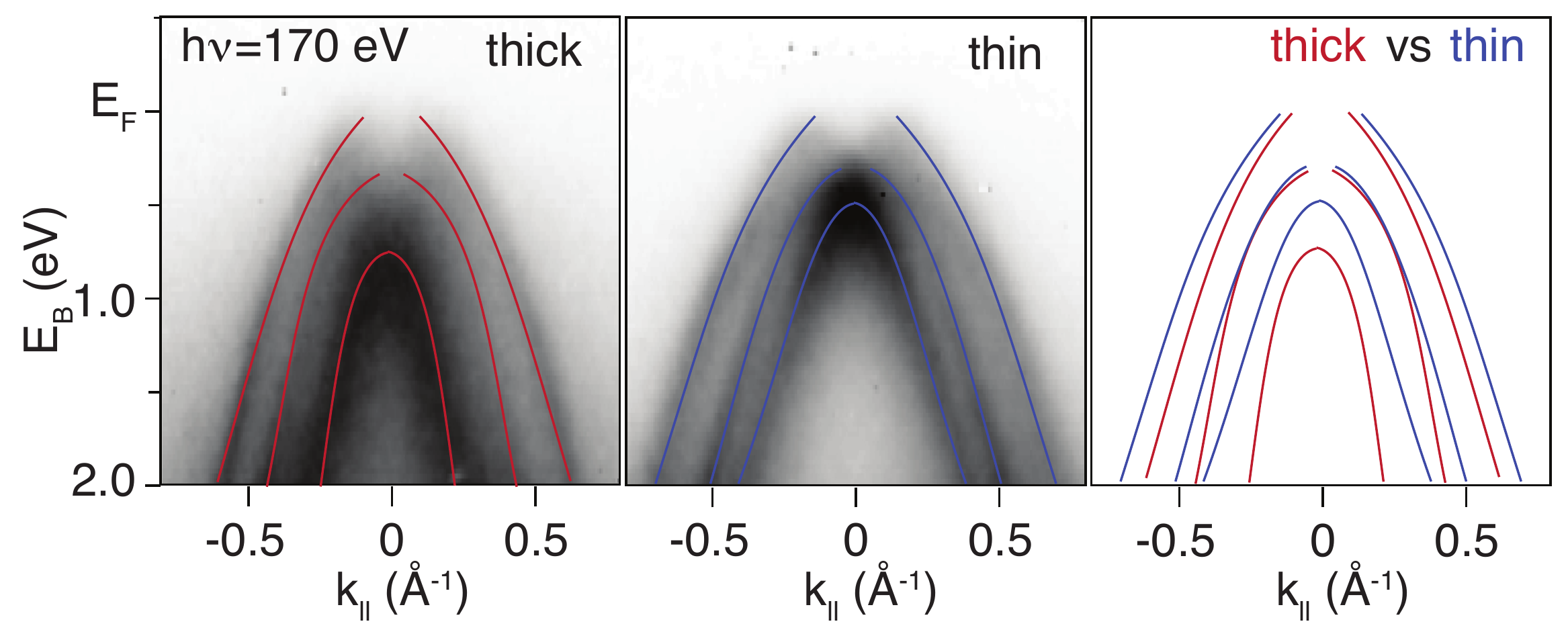}
	\caption{\textbf{Comparison of the band dispersions from a thin $\delta$-layer sample and a thick sample.} Measurements carried out at a photon energy of 170~eV (corresponding to a slice through the bulk BZ center). (a) the 3.0~$\mu$m bulk-like sample and (b) the 1.8~nm $\delta$-layer. Both figures are overlaid with parabolic schematic bands, as a ``guide-to-the-eye''. (c) comparison of the two sets of parabolic bands showing that there is a small, but significant, differnence in the effective mass for the two samples.}
	\label{fig:effectivemass}
\end{figure}

Dispersions in $E(k_\parallel)$ acquired with selected photon energies on both the $\delta$-layer and bulk film sample are presented in Fig.~\ref{fig:Ekpara}. While we cannot with complete certainty say that the dispersions are identical, the differences between the datasets are minor, and can be attributed to slight variations in doping concentration, sample alignment and impurities. Thus, in addition to not observing quantum well states in the $\delta$-layer, the thickness of the dopant layer does not appear to appear to alter the diamond occupied electronic structure significantly. On the other hand, measurements performed at lower photon energy (Fig.~\ref{fig:effectivemass}), and therefore with increased surface sensitivity, do appear to show a small change in the effective mass of the parabolic band maxima of the $\delta$-layer sample (compared to the thick film). This is accentuated in Fig.~\ref{fig:effectivemass}(c) in which the dispersions for both samples are plotted together. 

The observed modification of the effective mass can be attributed to electron correlations. Electron-electron correlations have been demonstrated to induce a bandwidth narrowing and to increase the effective mass of the charge carriers when the dimensionality of the sample is reduced \cite{Valla:2002,Qazilbash:2009,King:2014a}. This picture finds full agreement with our data where both a bandwidth narrowing and an increase of the hole effective mass is observed. In addition, electron-electron interactions are expected to be significant in stabilizing the electronic structure of superconductors and of materials which exhibit a metal-insulator transition, and both of these effects have been documented for boron-doped diamond \cite{Yokoya:2005,Klein:2007,Ekimov:2004,Calandra:2008}.  In these terms,  boron $\delta$-doped diamond would constitute a perfect playground for exploring the role of correlation effects and  putative high temperature superconductivity \cite{Kortus:2005}.  In any case, the fact that a modification of the effective mass can be seen adds assurance that a dense and narrow dopant profile is indeed present on the lengthscale probed by our ARPES measurements. 

Calculations indicate that, for a 1.8~nm boron doped $\delta$-layer in diamond, there will be occupied quantum well states located below the Fermi level \cite{Chicot:2014,Fiori:2010} and, thus, presumably observable in ARPES.  The lack of such states in our data  suggests that either our sample differs from the calculated systems, or that the calculations are an incomplete description of the physical system. 

On possible discrepancy between our measurements and the calculations is the layer thickness.  Our $\delta$-layer sample is grown so as to produce a dopant profile of 1.8~nm, however, the measured profile by SIMS is limited by the resolution of the instrument. It is therefore possible that the actual dopant profile is sharper, or slightly broader, then the nominal thickness.  As shown by Chicot~\textit{et al.}~reducing the thickness of the $\delta$-layer leads to a shallower and narrower potential and may create a situation where there are no longer occupied quantum well states (See Refs.\ \onlinecite{Chicot:2014,supp} for details). On the other hand, from the same calculations, a slightly broader profile is expected to still produce occupied confined states.


Another possible cause for discrepancy is the asymmetry of the confinement potential. In the work of Chicot~\textit{et al.}\ and Fiori~\textit{et al.}~the $\delta$-layer is either sandwiched between two 500~nm slabs of diamond (the ``infinite'' case), or sandwiched between a 500~nm diamond slab and a 25~nm diamond layer with a Schottky contact (the ``semi-infinite'' case). In our experiments, the $\delta$-layer is not encapsulated, and hence the potential gradient on the diamond/vacuum interface will be dissimilar (steeper) relative to the bulk side (see Ref.\ \onlinecite{supp} for details). This modification to the confinement potential is small, but in principle may cause the occupied states to shift further below the Fermi level (relative to a symmetric well). On the other hand, this is not expected to hinder detection by ARPES, and has not hindered comparable studies on unencapsulated SI:P $\delta$-layers\cite{Mazzola:2019}. 


It is also conceivable that the available calculations may not accurately describe the physical system. The calculations of Fiori~\textit{et al.}\ and Chicot~\textit{et al.}\ assumed that the relative dielectric constant within the $\delta$-layer is 5.7, that of undoped diamond. While the literature does not cover the specific case of doped diamond, it is known that the dielectric constant of highly doped semiconductors varies with dopant density.\cite{Ristic:2004,Mazzola:2019} The calculations of Fiori~\textit{et al.}\ and Chicot~\textit{et al.}\  also give no details of the screening potential used as part of the calculations. Correctly accounting for the dielectric constant and screening potential will naturally influence the calculations of the potential and charge distribution, and thus the predicted energies of the resulting confined states. Furthermore, Chicot~\textit{et al.}\ and Fiori~\textit{et al.}\ use isotropic effective masses ($m_{lh} = 0.303m_0, m_{hh} = 0.588m_0$, for the light-hole and heavy-hole states, respectively) taken from the work of Willatzen~\textit{et al.}\ in their calculations.\cite{Willatzen:1994} There is little agreement in the literature on hole effective masses in bulk diamond (See table~\ref{table:Emass}), and in highly doped $\delta$-layers the effective mass anyway appears to be modified. As energy eigenvalues calculated with the Schr{\"o}dinger equation are dependent on effective mass, it is straightforward to see that this choice will directly effect the theoretical description of the $\delta$-layer electronic structure. Furthermore, Fiori~\textit{et al.}\ and Chicot~\textit{et al.}\ do not address the positioning of the dopant atoms, in particular whether they have an ordered or disordered arrangement within the $\delta$-layer.  In the case of Si:P $\delta$ layers, ordering of the dopants influences the calculated electronic structure in the $\delta$-layer.\cite{Lee:2011b,Miwa:2014a} Such a study has not yet been published for boron-doped $\delta$-layers in diamond, but it is reasonable to infer that dopant order, or lack thereof, may also influence the $\delta$-layer electronic structure. In short, there are several differences in the representation of the $\delta$-layer in the calculations compared to the real sample, but none which are obviously responsible for the lack of quantised states in the occupied bandstructure.

\begin{table}
 	\caption{A selection of calculated (Linear Muffin Tin Orbital `LMTO' and Density Functional Theory `DFT') and measured effective masses for the light ($m_{lh}$) and heavy ($m_{hh}$) hole band in diamond based on published work; $m_0$ is the free electron mass.}
 	 \label{table:Emass}
 	 \begin{ruledtabular}
 	 	\begin{tabular}{p{3.4cm}p{1.3cm}p{1.3cm}p{2.4cm}}
 		 Citation & $m_{lh}$ & $m_{hh}$ & Method \\ 
          \hline
           		 Willatzen~\textit{et al.}\, [\onlinecite{Willatzen:1994}] & $0.303m_0$ & $0.588m_0$ & LMTO \\ 
          L{\"o}f{\aa}s~\textit{et al.}\, [\onlinecite{Lofas:2011}] & $0.309m_0$ & $0.600m_0$ & DFT (GW) \\ 
         Naka~\textit{et al.}\, [\onlinecite{Naka:2013}] & $0.263m_0$ & $0.653m_0$ & Cyclotron
 		 \end{tabular}
 	 \end{ruledtabular}
 \end{table}


It is interesting to notice that the electrical transport measurements of Chicot~\textit{et al.}\ also found no quantum enhancement of the hole mobility in encapsulated diamond $\delta$-layers,\cite{Chicot:2012,Chicot:2014} suggesting that there was no quantum confinement in this case either. Whilst it is possible that our ARPES investigation has somehow failed to observe quantised states, the implications from transport studies seem to support the notion that such states are simply not present. 

It is possible that the $\delta$-layer samples grown with current CVD processing methods\cite{Butler:2017,Vikharev:2016,Volpe:2012} are not able to produce a sufficiently sharp and narrow dopant profile, or that the lack of quantisation has another cause. In any case, further work from both a theoretical and experimental standpoint will be necessary to determine whether diamond $\delta$-layer samples can be fabricated in which quantised confined states are observable.



Finally, we consider the impact these findings have for the application of nanostructured boron-doped diamond grown using current CVD processing methods.  The growth of high quality diamond layers typically proceeds at rates on the order of micrometers per hour\cite{Spitsyn:1981},  with the diamond growth rate and defect density typically being inversely related. Thick, high-quality diamond films are therefore associated with long processing times and high cost.  While defect density is not necessarily a concern for all applications of doped diamond structures, properties such as electrical and thermal conductivity are hampered by high defect densities.\cite{McNamara-Rutledge:1995}  Our findings indicate that bulk-like electronic properties can be achieved in exceptionally thin films; which may be seen as an advantage when cost-effective high-quality fabrication is desired.  Whilst quantum effects certainly have their uses in quantum electronics,\cite{Ando:1982,Zwanenburg:2013,Bimberg:1999} the era of continued downscaling of traditional silicon-based devices is reaching its limit.  The persistence of bulk-like diamond electronic  properties at the single nanometer scale suggests that such limitations of downscaling are of less concern for diamond-based electrical components (such as piezoresistive diamond sensors\cite{Werner:1998a}), thus expanding the potential applications for nanoscale diamond components.

\section*{Conclusion}
This work uses ARPES to compare the occupied electronic structure of $\delta$-doped diamond with bulk-doped diamond, in order to determine if $\delta$-doped diamond grown with current CVD processing techniques produces a potential sufficiently strong and narrow to give rise to quantum confinement, as suggested by the Poisson-Schr{\"o}dinger calculations of Chicot~\textit{et al.}\ and Fiori~\textit{et al.} We instead observed that the occupied electronic structure of the $\delta$-doped layer was similar to that of a bulk-like doped film, with no additional features to attribute to electron-occupied quantum well states and no modification of the pre-existing bands except for a small modification of the effective mass. While this is contrary to the calculations, the electrical transport measurements performed by Chicot~\textit{et al.}\,\cite{Chicot:2012,Chicot:2014} also showed no quantum confinement related enhancement of the hole mobility in $\delta$-doped diamond, also suggesting that there may be no quantum confinement; further simulation studies will be necessary to understand the physical cause behind this. Based on our results it can be expected that regardless, of the size of doped diamond components, the desirable electronic properties of bulk doped diamond will be retained without being influenced by quantum confinement, an advantageous property for developing miniturised electrical components. \\

\noindent\textbf{Acknowledgements}
AKS and ACPT contributed equally to this  work. RBJ acknowledges the UK's Engineering and Physical Sciences Research Council (EPSRC) for partial funding for this activity (EP/H020055/1) as well as The EC's Horizon 2020 programme for support from the ``GREENDIAMOND'' project (ID: 640947). Oliver A. Williams and Souman Mandel (University of Cardiff, UK) are sincerely thanked for the growth of the thick, bulk-like doped diamond film used here. $\delta$-layer samples were supplied from a collaboration with Diamond Microwave Devices Ltd and Element Six Ltd, whilst the other was produced at the IAP RAS under the guidance of the authors JEB and AV, as part of the programme supported by Act 220 of the Russian Government (Agreement no. 14.B25.31.0021). Experimental results were acquired at the ADRESS beamline at SLS, the VUV beamline at Elettra and at beamline I4 of MAX-III. We acknowledge support from Marco Caputo and Vladimir Strocov, Balasubramanian Thiagarajan, and the other support staff at these facilities. This work was supported by the Research Council of Norway through its Centres of Excellence funding scheme, Project No. 262633, ``QuSpin'', and through the Fripro program, Project No. 250985 ``FunTopoMat''. This work was supported by the Danish Council for Independent Research, Natural Sciences under the Sapere Aude program (Grants No. DFF-4002-00029 and DFF- 6108-00409) and by VILLUM FONDEN via the Centre of Excellence for Dirac Materials (Grant No. 11744) and the Aarhus University Research Foundation.

\bibstyle{apsrev}
\bibliography{DiamondDeltaLayers}

\end{document}